\documentclass{article}
\usepackage{amsmath,graphicx}

\usepackage{amsmath,graphicx}
\usepackage{algorithm}
\usepackage{algpseudocode}
\usepackage{pifont}
\usepackage{amssymb}
\usepackage{bm}
\usepackage{multirow}
\usepackage{adjustbox}
\usepackage{tabularx}
\usepackage{tikz}
\usepackage{caption}
\usepackage{subcaption}
\usepackage{verbatim}
\usepackage{float}
\usepackage{commath}
\usepackage{url}
\usepackage{color, colortbl}

\usepackage[backend=bibtex,style=ieee,minbibnames=1,maxbibnames=4]{biblatex}
\usepackage{spconf}
\usepackage[keeplastbox]{flushend}
\newcommand{\Anurag}[1]{{\color{blue}{\bf Anurag: }\em{#1}}}

\definecolor{LightCyan}{rgb}{0.88,1,1}

\title{TPARN: Triple-path attentive recurrent network for time-domain multichannel speech enhancement}

\name{Ashutosh Pandey$^{\textup{1}*}$, Buye Xu$^\textup{1}$, Anurag Kumar$^\textup{1}$, Jacob Donley$^\textup{1}$, Paul Calamia$^\textup{1}$ and DeLiang Wang$^\textup{2}$\thanks{$^{*}$Work done during internship at Facebook Reality Labs Research.}}
\address{$^\textup{1}$Facebook Reality Labs Research, USA\\
$^\textup{2}$Department of Computer Science and Engineering, The Ohio State University, USA}
\begin{document}
\ninept
\maketitle
\begin{abstract}
In this work, we propose a new model called triple-path attentive recurrent network (TPARN) for multichannel speech enhancement in the time domain. TPARN extends a single-channel dual-path network to a multichannel network by adding a third path along the spatial dimension. First, TPARN processes speech signals from all channels independently using a dual-path attentive recurrent network (ARN), which is a recurrent neural network (RNN) augmented with self-attention. Next, an ARN is introduced along the spatial dimension for spatial context aggregation. TPARN is designed as a multiple-input and multiple-output architecture to enhance all input channels simultaneously. Experimental results demonstrate the superiority of TPARN over existing state-of-the-art approaches.
\end{abstract}
\section{Introduction}
\label{sec:intro}

Speech enhancement is concerned with improving the intelligibility and quality of a speech signal degraded by noise and reverberation. The most basic approach to speech enhancement is monaural processing, where recordings from a single microphone are utilized \cite{loizou2013speech}. Single-channel methods can obtain good enhancement, but are limited in capability to utilize only time-frequency (T-F) information. Multichannel speech enhancement aims at utilizing both T-F and spatial information by using recordings from multiple microphones \cite{benesty2008microphone, gannot2017consolidated}.

Supervised speech enhancement using deep neural networks (DNNs) represents the mainstream methodology for speech enhancement \cite{wang2017supervised}. For multichannel processing, a popular approach is to incorporate DNNs with traditional spatial filters, such as an MVDR beamformer \cite{erdogan2016improved, heymann2016neural}. A DNN is first used to estimate second-order statistics of speech and noise which are then used for computing beamformer weights. Another approach is to train a DNN with spatial features, such as inter-channel time, phase or level difference \cite{wang2018combining, wang2018multi}. A more recent trend is to use end-to-end supervised learning, where spatial information becomes an implicit part of supervised learning \cite{tolooshams2020channel, wang2020multi}. Wang et al. \cite{wang2020multi} proposed a dense convolutional recurrent network (DCRN) for multi-microphone complex spectral mapping, where real and imaginary components of the clean spectrum are directly predicted from the multichannel noisy spectrum. In \cite{tolooshams2020channel}, authors used inspiration from  complex beamforming to propose a novel channel-attention mechanism inside a dense UNet. 

Moreover, time-domain speech enhancement using DNNs has also gained considerable attention in recent years \cite{Pandey2018, luo2019conv, pandey2021dense, luo2020dual}. Time domain networks directly map noisy speech samples to clean speech samples, and as a result, feature extraction becomes an implicit part of the learning process. Even though highly effective in removing additive interference, time-domain approaches have not yet been established for removing room reverberation, a convolutive interference \cite{luo2018real}. Time-domain networks have also been explored for end-to-end multichannel speech enhancement \cite{tawara2019multi, liu2020multichannel, luo2020end}, however reported  performances are far from satisfactory. 

In this work, we propose a novel approach for end-to-end time-domain multichannel speech enhancement. We refer to it as \emph{TPARN: Triple-path Attentive Recurrent Network}.
 The key idea in the TPARN design is to extend a dual-path network
 \cite{ luo2020dual} with a third path along the spatial dimension. The audio input signals from all channels are first divided into short chunks which are then processed by the TPARN system in three stages. These three stages for processing include: \emph{intra-chunk processing} for local temporal modeling, \emph{inter-chunk processing} for global temporal modeling, and \emph{inter-channel processing} for spatial modeling. 
 
 Intra-chunk and inter-chunk processing are performed independently for all the channels using a dual-path attentive recurrent networks (ARN) \cite{pandey2020dual}, which are RNNs augmented with self-attention \cite{merity2019single, pandey2021self}. A combination of RNN and self-attention has been proven to be effective for speech processing tasks \cite{pandey2020dual, tan2020sagrnn, pandey2021self, chen2020dual}.  The inter-channel processing introduced by us can be modeled using different methods and we explore RNN, self-attention network, and ARN along the spatial dimension. We find ARN to be slightly superior compared to the other two. Besides, with an explicit capability to capture spatial information through neural network based inter-channel processing, our TPARN framework has additional desirable characteristics.  For example, TPARN is designed as a multiple-input and multiple-output (MIMO) architecture to enhance all input channels simultaneously. These multi-channel outputs can be further processed by a downstream system if needed.  We train and evaluate TPARN on two different datasets with varying degrees of reverberation and noise. We show that TPARN obtains better results than state-of-the-art approaches on both datasets. We also find performance improvements to be more significant on a  difficult dataset. 

\section{Model Description}
\subsection{Problem Definition}
A multichannel noisy signal $\bm{X} = \{\bm{x}_{1}, \dots, \bm{x}_{P}\}\in \mathbb{R}^{P \times N}$, where $P$ is the number of microphones and $N$ is the number of samples, is modeled as
\begin{equation}
\begin{split}
x_{p}(n) &= y_{p}(n) + z_{p}(n) \\
               &= d_{p}(n) + r_{p}(n) + z_{p}(n) \\
               & = d_{p}(n) + u_{p}(n)
\end{split} 
\end{equation}
where $p = 1, 2 \dots P$ and $n = 0, 1, \dots N-1$. $\bm{x}_{p}$ represents noisy signal at microphone $p$.  $\bm{y}$ is the received speech including direct-path speech $\bm{d}$ and reverberated speech $\bm{r}$. $\bm{z}$ is the noise and $\bm{u}$ is the overall interference including noise and reverberation. The goal of multichannel speech enhancement is to get a close estimate $\bm{\hat{d}}_{r}$ of the direct-path clean speech at a reference microphone $r$ from $\bm{X}$.
\begin{figure}[!b]
\centering
\includegraphics[width=\columnwidth, keepaspectratio]{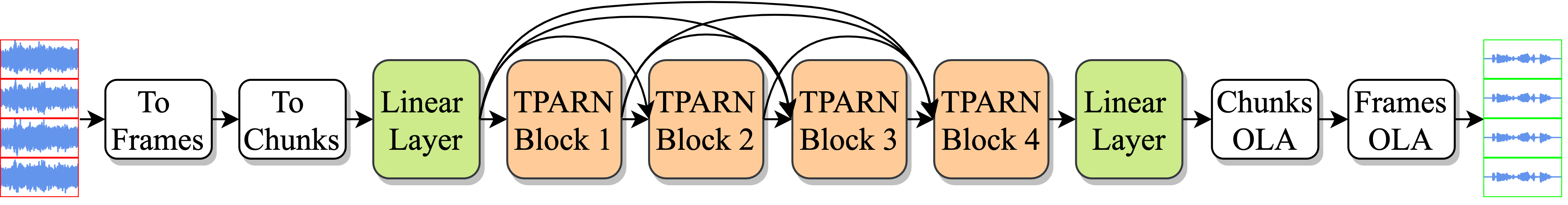}
\caption{\it The proposed TPARN architecture for multichannel speech enhancement.}
\label{fig:tparn}
\end{figure}
\subsection{Triple-path Attentive Recurrent Network}
The overall schema of the proposed TPARN architecture is shown in Fig. \ref{fig:tparn}. It consists of an input linear layer, four TPARN blocks, and an output linear layer. 

\begin{figure}[!h]
\centering
\includegraphics[width=0.94\columnwidth, keepaspectratio]{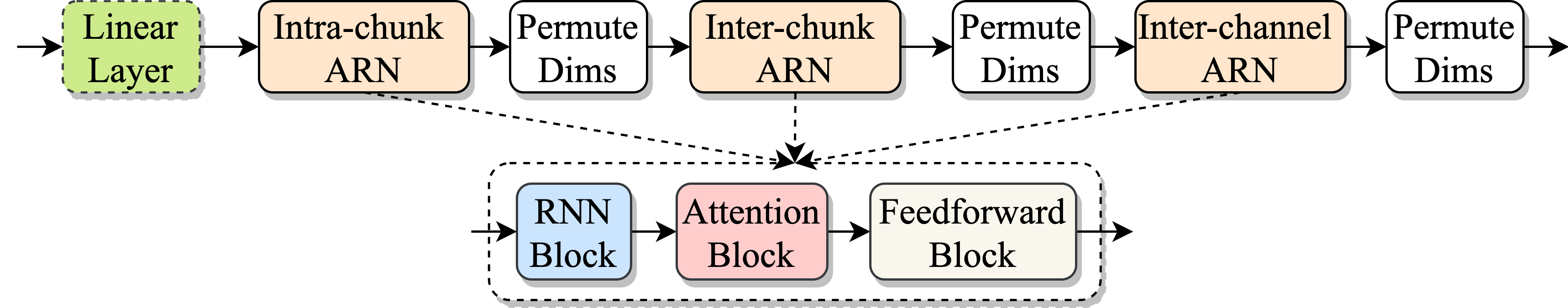}
\caption{\it TPARN block.}
\label{fig:tparn_block}
\end{figure}

An input signal $\bm{X} \in \mathbb{R}^{P \times N}$ is first converted into frames, $\mathcal{T} = [\bm{X}_{1}, \ldots, \bm{X}_{T}] \in \mathbb{R}^{P \times T \times L}$, using a frame size of $L$ samples and frame shift of $K$. $T$ is the total number of frames. The frames in $\bm{\text{T}}$ are arranged into chunks with a chunk size of $R$ and chunk shift of $S$, leading the input being represented as $ \mathcal{C}= [\bm{\text{C}}_{1}, \ldots \bm{\text{C}}_{C}] \in \mathbb{R}^{P \times C \times R \times L}$, where $C$ is the number of chunks. Next, the frames of size $L$ in $\mathcal{C}$ are projected to $D$ dimensions using the input linear layer, which are then processed by a stack of 4 TPARN blocks. The architecture of a TPARN block is shown in Fig. \ref{fig:tparn_block}. The TPARN blocks are densely connected. The input to TPARN blocks are 4d tensors of shape $P\times C \times R \times k \cdot D$, which are obtained by concatenating the output from the linear layer encoder and the outputs from preceding TPARN blocks. $k \in \{1, 2, 3, 4\}$ denotes block id. 

 For $k > 1$, a linear layer is used at the input to project features of size $k \cdot D$ to $D$. Within a TPARN block, the inputs are processed using a stack of three ARNs: intra-chunk ARN, inter-chunk ARN and inter-channel ARN. The intra-chunk ARN operates independently over all chunks by rearranging its input to shape $P\cdot C \times R \times D$, and using an ARN that treats the first, second and third dimensions as batch, sequence and feature dimensions respectively. Similarly, the inter-chunk ARN combines all chunks together by rearranging its input to shape $P\cdot R \times C\times D$. The inter-channel ARN operates along the channel dimension (spatial dimension) by rearranging its input to shape $R\cdot C \times P \times D$. The sequence length and the batch size for an utterance for different ARNs are given in Table 1. 

\begin{figure}[!b]
\centering
\includegraphics[width=0.8\columnwidth, keepaspectratio]{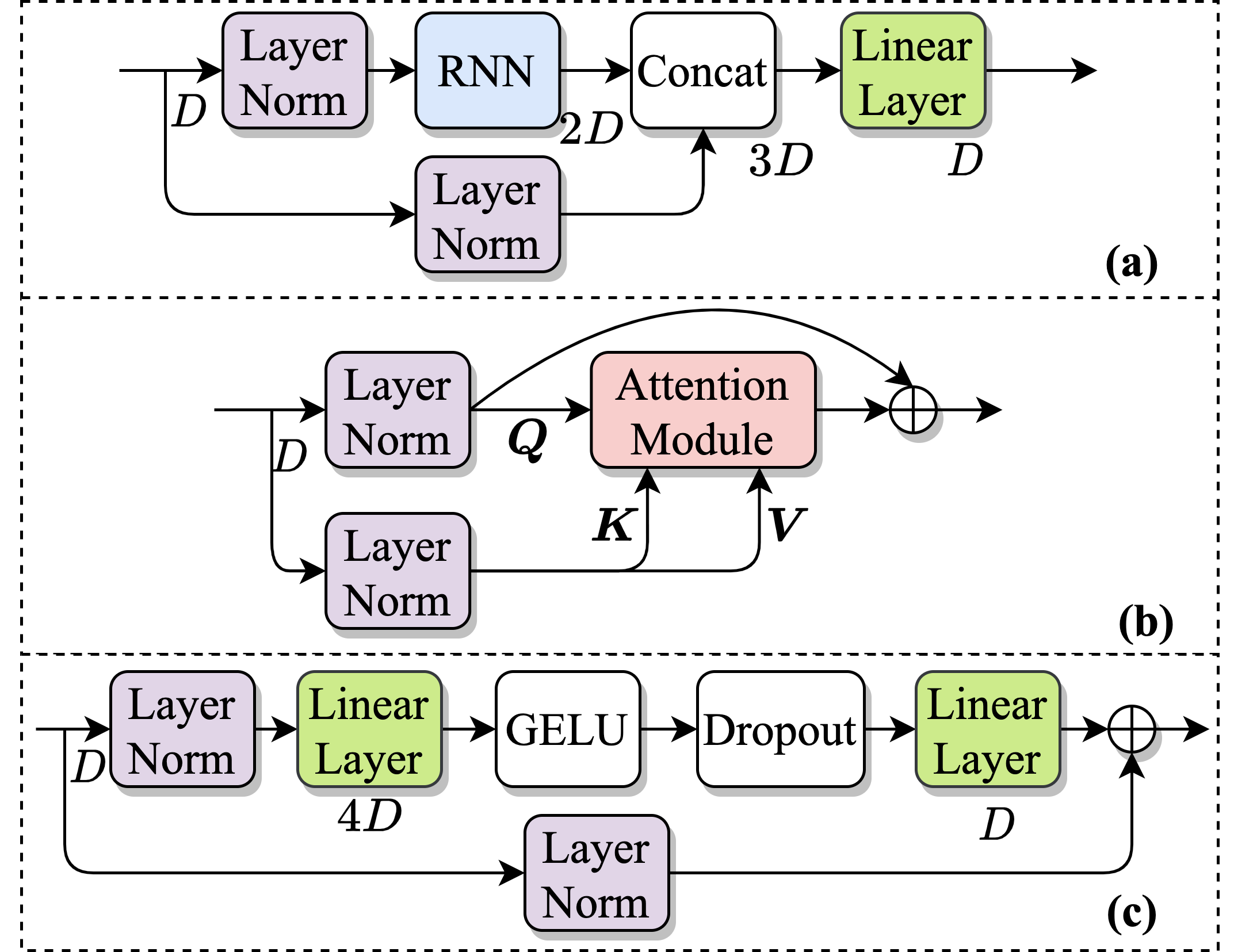}
\caption{\it (a) RNN block,  (b) Attention block, (c) Feedforward block.}
\label{fig:rnnt}
\end{figure}

 Fig. 2 also depicts the architecture of an ARN which comprises of a stack of three blocks, RNN block, attention block, and feedforward block. The architecture of the RNN block, the attention block and the feedforward block is shown in Fig. 3. The input to RNN block is normalized using two independent layer normalization layers. We use separate normalization to make sure that the network has a capability to scale a given signal differently at different locations inside the network. The first layer-normalized input is fed to an RNN of hidden size $D$.  The output of the RNN is concatenated with the second layer-normalized input projected to size $D$ using a linear layer. 
 
\begin{figure}[!t]
\centering
\includegraphics[width=0.75\columnwidth, keepaspectratio]{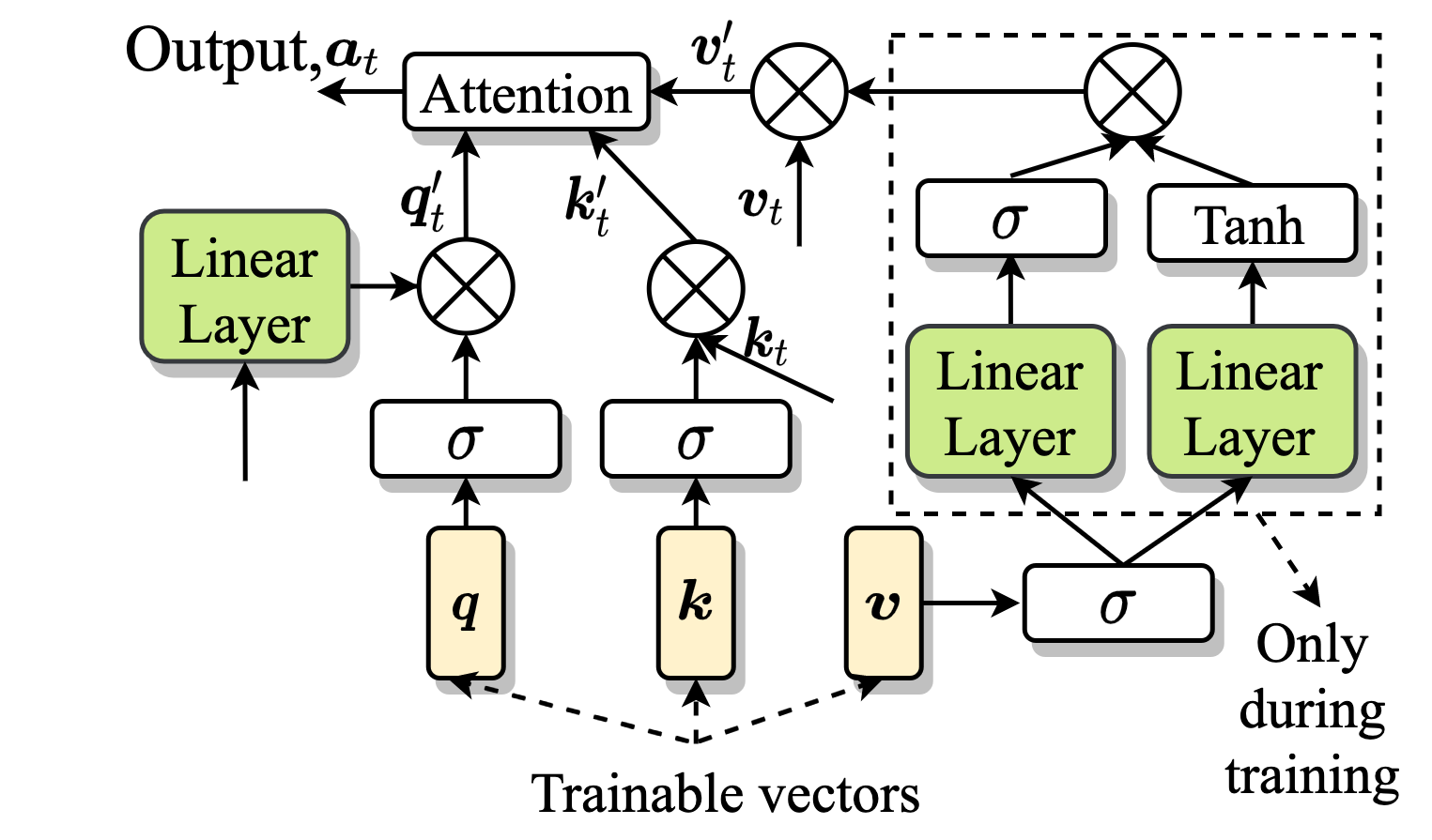}
\caption{\it Self-attention mechanism.}
\label{fig:attention_block}
\end{figure}

 The input to the attention block is layer-normalized using two independent layer-normalization layers. The first layer-normalized input is used as query, $\bm{Q}$, and the second layer-normalized input is used as key $\bm{K}$ and value $\bm{V}$ for a following attention module. The attention mechanism of the attention module, shown in Fig. \ref{fig:attention_block}, is borrowed from \cite{merity2019single} where its effectiveness with RNN for natural language processing tasks has been demonstrated. Its effectiveness for speech enhancement has also been established in \cite{pandey2020dual, pandey2021self}. 
 
 \begin{table}
\centering
\caption{ \it Input size to different ARNs for an utterance with $C$ chunks.}
\begin{adjustbox}{width=0.66\columnwidth}
\begin{tabular}{|c|c|c|c|}
\cline{2-4}
\multicolumn{1}{c|}{}& Batch Size & Seq. Length & Feature Size \\
\hline
Intra-chunk ARN & $P \cdot C$ & $R$ & $D$ \\
\hline
Inter-chunk ARN & $P \cdot R$ & $C$ & $D$ \\
\hline
Inter-channel ARN & $R \cdot C$ & $P$ & $D$ \\
\hline
\end{tabular}
\end{adjustbox}
\end{table}
 
 The attention module comprises three trainable vectors\\ \scalebox{1.0}{$\{\bm{Q}^{\prime}, \bm{K}^{\prime}, \bm{V}^{\prime}\} \in \mathbb{R}^{1 \times D}$}, and its inputs are $\{\bm{Q}, \bm{K}, \bm{V}\} \in \mathbb{R}^{B \times U \times D}$, where $B$ is the batch size and $U$ is the sequence length (see Table 1). $\bm{Q}, \bm{K}$, and $\bm{V}$ are refined using a gating mechanism given in the following equation. 

\begin{equation}
\begin{split}
\bm{K}_{r} &= \bm{K} \odot \text{Sigm}(\bm{K}^{\prime})\\
\bm{Q}_{r} &= \text{Lin}(\bm{Q}) \odot  \text{Sigm}(\bm{Q}^{\prime}) \\
\bm{V}_{r} &= \bm{V}\odot  [\text{Sigm}(\text{Lin}(\bm{V}^{\prime})) \odot \text{Tanh}(\text{Lin}(\bm{V}^{\prime}))]
\end{split}
\end{equation}  
where $\text{Sigm()}$ is the sigmoidal nonlinearity, $\text{Lin()}$ is a linear layer, and $\odot$ denotes elementwise multiplication. $\bm{Q}^{\prime}, \bm{K}^{\prime}$, and $\bm{V}^{\prime}$ are broadcast to match the shape of $\bm{Q}, \bm{K}$, and $\bm{V}$. Note that $\text{Sigm}(\text{Lin}(\bm{V}^{\prime})) \odot \text{Tanh}(\text{Lin}(\bm{V}^{\prime}))$ is a deterministic vector, and hence this operation is used only during training to better optimize  $\bm{V}^{\prime}$, and its final value is stored as a vector to use during evaluation.
 
 The final output of the attention block is computed as
 \begin{equation}
 \bm{A} = \text{Softmax}(\frac{\bm{Q}_{r} \bm{K}_{r}^{T}}{\sqrt{D}})\bm{V}_{r}
 \end{equation}

The input to the feedforward block in ARN is layer-normalized independently using two different layer normalization layers. The first layer-normalized input is projected to size $4D$ using a linear layer followed by  Gaussian error linear unit (GELU) nonlinearity and a dropout with dropout rate $D_r$. Finally, it is projected back to size $D$ and added to the second layer-normalized input. Note that we use dense connection in the RNN block and residual connections in the attention and the feedforward block.

The output of the final TPARN block is projected to size $L$ using the output linear layer. Next, chunks are combined together using chunk overlap-and-add (OLA) and then frames are combined together using frame OLA to get an enhanced multichannel waveform. Note that TPARN is a MIMO architecture that enhances all input channels simultaneously.

\subsection{Loss Function}
We use a recently proposed phase constrained magnitude (PCM) loss \cite{pandey2020densely} for training. The PCM loss was proposed to overcome an existing artifact issue with magnitude-based losses for time-domain networks. First, we compute an estimate of overall interference as
\begin{equation}
    \bm{\hat{U}} = \bm{X} - \bm{\hat{D}}
\end{equation}
Then, a magnitude-based loss is used between reference and estimated speech and noise.
\begin{equation}
L_{PCM}(\bm{D}, \bm{\widehat{D}}) = \frac{1}{2}\cdot L_{SM}(\bm{D}, \bm{\widehat{D}}) + \frac{1}{2} \cdot L_{SM}(\bm{U}, \bm{\widehat{U}})
\end{equation}
$L_{SM}$ is defined as \\
\begin{equation}
\begin{aligned}
L_{SM}(\bm{D}, \bm{\widehat{D}}) = \frac{1}{P \cdot T\cdot F}\sum_{p=0}^{P} \sum_{t=1}^{T}\sum_{f=1}^{F}&|(|\mathcal{D}_{p}^{r}(t, f)| + |\mathcal{D}_{p}^{i}(t,f)|)\\-&\ (|\widehat{\mathcal{D}}_{p}^{r}(t,f)| + |\widehat{\mathcal{D}}_{p}^{i}(t, f)|)|
\end{aligned}
\end{equation}

where $\bm{\mathcal{D}}_{p}$ is STFT of $\bm{d}_{p}$ and $\bm{\mathcal{D}}_{p}^{r}$ and $\bm{\mathcal{D}}_{p}^{i}$ respectively represent the real and the imaginary component of $\bm{D}_{p}$. $T$ is the number of frames and $F$ is the number of frequency bins.

\section{Experiments}
\subsection{Datasets}
We use two datasets for experiments. The first dataset is created using speech from WSJCAM0 corpus \cite{robinson1995wsjcamo} and noises from the REVERB challenge \cite{kinoshita2016summary}. This dataset uses a 4 microphones circular array with a radius of 10 cm, T60 in the range [0.2, 1.2] seconds, and direct speech-to-noise-ratio (SNR) in the range [5, 20] dB. More details about this dataset can be found in \cite{wang2020multi}. 

We also create a more challenging dataset using speech and noises from the DNS challenge 2020 corpus\footnote{\footnotesize{\url{https://github.com/microsoft/DNS-Challenge/blob/master/LICENSE}}}~\cite{reddy2020interspeech}. We select all speakers with one chapter in the dataset and randomly select 90\% of speakers for training, 5\% for validation, and 5\% for evaluation. After this, for each utterance a random chunk of a randomly sampled length with an activity threshold (from script in \cite{reddy2020interspeech}) greater than 0.6 is extracted. The length of utterances are sampled from [3, 6] seconds for training and [3, 10] seconds for test and validation. This results in a total of 53 k utterances for training, 2.6 k for validation, and 3.3 k for test. Next, all the noises from the DNS corpus are randomly divided into training, validation and test noises in a proportion similar to the number of speech utterances.

The algorithm to generate spatialized multichannel data from DNS speech and noises is given in Algorithm 1. All the points inside a room are sampled at least 0.5 m away from walls, and the distance between array center and different sound sources is kept between 0.75 m and 2 m. We use Pyroomacoustics \cite{scheibler2018pyroomacoustics} with hybrid approach where the image method with order 6 is used to model early reflections and ray-tracing is used to model the late reverberation. A similar approach to data generation was used in \cite{kabeli2021online}. 

\begin{table}[!t]
\centering
\caption{Comparison of Loss Functions for TPARN.}
\begin{adjustbox}{width=0.4\columnwidth}
\begin{tabular}{|c|c|c|c|}
\hline
& SI-SDR & \ STOI \  & \ PESQ \ \\
\hline
\hline
unproc. & -3.8 & 70.9 & 1.63 \\
\hline
\hline
MSE & 10.1 & 95.7 & 3.07 \\
\hline
PCM & \textbf{10.4} & \textbf{96.9} & \textbf{3.42} \\
\hline
\end{tabular}
\end{adjustbox}
\label{tab:loss_funcs}
\end{table}

\setlength{\textfloatsep}{0.4cm}
\begin{algorithm}[!t]
\caption{DNS dataset spatialization process.}
\begin{algorithmic}
\footnotesize
\For {\emph{split} in \{train, test, validation \}}
\For {speech utterances in \emph{split} }
\begin{itemize}
\item Draw room length and width from [5,10] m, and height from [3, 4] m;
\item Draw 1 array location and 1 speech source location;
\item Get 4 uniformly placed mic locations on a circle of radius 10 cm centered at array location;
\item Draw $N_{ns}$, number of noise sources, from [5, 10]
\item Draw $N_{ns}$ noise locations in room
\item Generate RIRs corresponding to speech source location and $N_{ns}$ noise locations for mic locations in circular array
\item Draw $N_{ns}$ noise utterances from noises in \emph{split} 
\item Propagate speech and noise signals to mics by convolving with corresponding RIRs
\item Draw a value $snr$ from [-10, 10] dB, and add speech and noises at each mic using a scale so that the overall direct speech SNR is $snr$;
\end{itemize}
\EndFor
\EndFor
\end{algorithmic}
\end{algorithm} 
\setlength{\floatsep}{0.4cm}

\subsection{Experimental settings}
All the utterances are resampled to 16 kHz.  We use $P = 4$, $L = 16$, $K = 8$, $R = 126$, $S = 63$, and $D = 128$ . For RNN, we use bidirectional long short-term memory networks (BLSTMs) with hidden size $D$ in each direction. Dropout rate $D_{r}$ in feedforward blocks is set to $5$\%.  We use phase constrained magnitude (PCM) loss in Eq. (6) for training TPARN. All the models are trained for 100 epochs on 4 second long utterances randomly extracted from training samples during training. A batch size of $8$ is used. Automatic mixed precision training is utilized for efficient training \cite{micikevicius2017mixed}. Learning rate is initialized with $0.0004$ and is dynamically scaled to half if the best validation score does not improve in five consecutive epochs.

We compare TPARN with two recently proposed complex spectrum based end-to-end models; DCRN \cite{wang2020multi} and channel-attention dense UNet (CA-DUNet)\cite{tolooshams2020channel} . We also compare it with two time-domain end-to-end models; residual speech denoising fully convolutional network (rSDFCN) \cite{liu2020multichannel}, and filter-and-sum network with transform average and concatenate module (Fasnet TAC) \cite{luo2020end}.

All the models are compared using three enhancement objective metrics: short-time objective intelligibility (STOI) \cite{taal2010short}, perceptual evaluation of speech quality (PESQ) \cite{rix2001perceptual}, and scale-invariant signal-to-distortion ratio (SI-SDR). STOI scores are reported in percentage. 


\subsection{Experimental results}
 We begin by comparing time-domain mean squared error (MSE) loss with spectral magnitude based PCM loss in Eq. (5). Results on WSJCAM0 dataset using TPARN are reported in Table \ref{tab:loss_funcs}. We see that even though MSE loss obtains good SI-SDR and STOI scores, it is considerably worse in terms of PESQ. This also suggests that PCM loss obtains bigger improvements for joint denoising and dereverberation compared to denoising in \cite{pandey2021dense}. We use PCM loss for the rest of the experiments with TPARN. 

Next, we explore the behavior of spatial  (inter-channel) ARNs at different locations inside a TPARN bloc: \emph{Pre} - before intra-chunk and inter-chunk ARN, \emph{Mid} -   between intra-chunk and inter-chunk ARN, and \emph{Post} - after inter-chunk and intra-chunk ARN. We also experiment with three different configurations for spatial processing: \emph{Attention} - removing the RNN block from ARN, \emph{RNN} - removing the attention block from ARN, and  \emph{ARN}. The results for different models are summarized in Table \ref{tab:spatial_location}. We see that for the WSJCAM0 dataset, different configurations of spatial ARN in TPARN block obtain similar results. However, for the challenging DNS dataset, \emph{RNN} is considerably worse in comparison of \emph{Attention} and \emph{ARN}. Moreover, best scores are obtained at \emph{Pre} location for \emph{Attention} and at \emph{Post} location for \emph{ARN}. These results indicate that RNNs are good for spatial modeling but may not suffice in extreme conditions, as in the DNS dataset. 

Additionally, we perform an ablation experiment (not reported here) on the number of spatial ARNs inside TAPRN by removing spatial ARNs from TPARN blocks at different locations. We find that a TPARN with 3 spatial ARNs in the first, second and the fourth TPARN block obtains consistently better results for both datasets and for different learning strategies, such as multiple-input and single-output (MISO) and MIMO. 

Finally, we compare best TPARN results with baseline models in Table 4. We report single-input and single-output (SISO), MISO, and MIMO results for DCRN and TPARN, and MISO results for the rest of the models, as in their original studies. TPARN MIMO is converted to MISO by averaging the output of the final TPARN block. We notice a very interesting observation that TPARN-SISO is worse than DCRN-SISO, but TPARN-MIMO and TPARN-MISO are better than DCRN-MISO and DCRN-MIMO. This suggests that TPARN exploits spatial information to a larger extent than in DCRN. Further, DCRN-MIMO is worse than DCRN-MISO, but TPARN-MIMO is slightly better than TPARN-MISO. This indicates that TPARN is capable of MIMO learning without any performance degradation, which provides additional advantage of enhancing all channels simultaneously with spatial cue preservation. Moreover, performance improvements over the baseline models are even better for the difficult DNS dataset. For example, TPARN is better by $3.8$ dB in SI-SDR, $1.9$\% in STOI and $0.12$ in PESQ compared to the second best DCRN-MISO. An exception in baseline models is CA-DUNet that obtains impressive SI-SDR but drastically worse PESQ for WSJCAM0. 

\begin{table}[t]
\centering
\caption{\it Comparisons between different spatial processing modules at different locations in TPARN blocks. }
\begin{adjustbox}{width=.9\columnwidth}
\begin{tabular}{|c||c||c|c|c||c|c|c|}
\cline{2-8}
\multicolumn{1}{c|}{ }& Test Dataset & \multicolumn{3}{c||}{WSJCAM0} & \multicolumn{3}{c|}{DNS} \\
\cline{2-8}
\multicolumn{1}{c|}{ }& Test Metric & SI-SDR & STOI & PESQ & SI-SDR & STOI & PESQ \\
\hline
Unproc. & Loc. $\downarrow$& -3.8 & 70.9 & 1.38 & -7.6 & 63.8 & 1.38 \\
\hline
\hline
\multirow{3}{*}{ \emph{Attention}} & \emph{Pre} & 10.3 & 96.8 & 3.40 & 7.6 & \textbf{91.1} & \textbf{2.66} \\
& \emph{Mid} & 10.1 & 96.8 & 3.39 & 7.0 & 90.2 & 2.56 \\
& \emph{Post} & 10.2 & 96.8 & 3.40 & 6.9 & 90.2 & 2.56 \\
\hline
\hline
\multirow{3}{*}{ \emph{RNN} } & \emph{Pre} & 10.3 & 96.8 & 3.40 & 6.9 & 90.0 & 2.55 \\
& \emph{Mid} & 10.3 & 96.9 & 3.44 & 7.1 & 90.4 & 2.59 \\
& \emph{Post} & 10.3 & 96.9 & 3.41 & 7.2 & 90.3 & 2.57 \\
\hline
\hline
\multirow{3}{*}{ \emph{ARN} } & \emph{Pre} & 10.4 & 96.8 & 3.40 & 7.4 & 90.5 & 2.59 \\
& \emph{Mid} & 10.4 & 96.9 & 3.41 & 7.1 & 90.4 & 2.59 \\
& \emph{Post} & \textbf{10.4} & \textbf{96.9} & \textbf{3.42} & \textbf{7.8} & \textbf{91.1} & 2.65 \\
\hline
\end{tabular}
\end{adjustbox}
\label{tab:spatial_location}
\end{table}

\begin{table}[!t]
\centering
\caption{ \it Comparisons with baseline models. WM: waveform mapping (time-domain), CRM: complex ratio masking. CSM: complex spectral mapping.}
\begin{adjustbox}{width=0.9\columnwidth}
\begin{tabular}{|c||c||c|c|c||c|c|c|}
\cline{2-8}
\multicolumn{1}{c||}{}& Test Dataset & \multicolumn{3}{c||}{WSJCAM0} & \multicolumn{3}{c|}{DNS} \\
\cline{2-8}
\multicolumn{1}{c||}{}& Test Metric & SI-SDR & STOI & PESQ & SI-SDR & STOI & PESQ \\
\cline{1-8}
Approach $\downarrow$& Unprocessed & -3.8 & 70.9 & 1.38 & -7.6 & 63.8 & 1.38 \\
\hline
\hline
WM & rSDFCN-MISO \cite{liu2020multichannel} & 4.2 & 83.8 & 2.00 & -2.2 & 68.5 & 1.49 \\
CRM & CA-DUNet-MISO \cite{tolooshams2020channel}& \textbf{10.7} & 96.0 & 2.88 & 3.5 & 83.3 & 1.99 \\
WM & Fasnet TAC-MISO \cite{luo2020end} & 8.2 & 94.7 & 2.93 & 4.7 & 86.5 & 2.26 \\
CSM & DCRN-SISO \cite{wang2020multi} & 6.6 & 93.6 & 2.90 & 3.9 & 89.8 & 2.60 \\
CSM & DCRN-MISO \cite{wang2020multi} & 9.4 & 96.5 & 3.31 & 4.6 & 90.1 & 2.57 \\
CSM & DCRN-MIMO \cite{wang2020multi} & 8 .0& 95.9 & 3.27 & 3.6 & 89.4 & 2.57 \\
\hline
\hline
WM & TPARN-SISO & 5.1 & 93.6 & 2.92 & 3.0 & 84.1 & 2.14 \\
WM& TPARN-MISO & 10.2 & 96.8 & 3.40 & 8.2 & 91.6 & 2.72 \\
WM& TPARN-MIMO &  10.4 & \textbf{96.9} & \textbf{3.43} &  \textbf{8.4} &\textbf{91.9} &\textbf{2.75} \\
\hline
\end{tabular}

\end{adjustbox}
\end{table}

\vspace{-11pt}
\section{Conclusions}
We have proposed a novel triple-path attentive recurrent network for multichannel speech enhancement in the time-domain. TPARN is designed as a simple extension of a single-channel dual-path network to multichannel network by adding a third-path along the spatial dimension. TPARN is a multiple-input and multiple-output (MIMO) architecture that can simultaneously enhance signals at all microphones. We have shown that TPARN obtains significantly better results than other state-of-the-art models in very noisy and reverberant conditions. Future research includes exploring TPARN for ad-hoc array processing and moving sources.



\setlength\bibitemsep{0.5em}
\atColsBreak{\vskip0.5em}
\printbibliography

\end{document}